# Separation of cardiac and respiratory components from the electrical bio-impedance signal using PCA and fast ICA

Yar M. Mughal [#1], A. Krivoshei [#1 *2], P. Annus [#1 *2]

[#]Thomas Johann Seebeck, Department of Electronics, TUT
Ehitajate tee 5, 19086 Tallinn, Estonia
[1]yar@elin.ttu.ee
[2]andreik@elin.ttu.ee
[3]paul.annus@elin.ttu.ee

*E LIKO Competence Centre in Electronics, Info- and Communication Technologies
Mäealuse 2/1, 3rd floor, 12618 Tallinn, Estonia.

*Abstract*— This paper is an attempt to separate cardiac and respiratory signals from an electrical bio-impedance (EBI) dataset. For this two well-known algorithms, namely Principal Component Analysis (PCA) and Independent Component Analysis (ICA), were used to accomplish the task. The ability of the PCA and the ICA methods first reduces the dimension and attempt to separate the useful components of the EBI, the cardiac and respiratory ones accordingly. It was investigated with an assumption, that no motion artefacts are present. To carry out this procedure the two channel complex EBI measurements were provided using classical Kelvin type four electrode configurations for the each complex channel. Thus four real signals were used as inputs for the PCA and fast ICA. The results showed, that neither PCA nor ICA nor combination of them can not accurately separate the components at least are used only two complex (four real valued) input components.

*Keywords*— Principal Component Analysis, fast Independent Component Analysis, Electrical Bio-impedance, Cardiac Signal and Respiratory signal

## I. Introduction

The Blind Source Separation (BSS) is forth-coming field of interest to separate useful components in the case of using the multi-source signals. The BSS is an approach which allows estimating original statistically independent sources from only observed mixture of these sources without any known *a priori* information.

The BSS can be found in many applications, such as biomedical, telecommunication, image and speech signal processing [1-4]. Many algorithms are used to implement the BSS. They are based on a statistical independence property of the separated signals [3, 4].

The Principal Component Analysis (PCA) algorithm is a useful technique to uncorrelate the components, but the Independent Component Analysis (ICA) algorithm as an implementation of the BSS; it is suitable to separate the original independent components [1-4].

The purpose for combining the both algorithms (PCA and ICA) is to decrease the dimension of the electrical bio-impedance (EBI) data set before implementing the ICA algorithm, hence to make easier for the ICA to separate the cardiac and respiratory components.

In this study, it was tried to apply the Principal Component Analysis (PCA) and fast Independent Component Analysis (ICA) algorithms to the electrical bio-impedance (EBI) dataset, which consist of two simultaneously measured complex EBI signals containing both the respiratory and cardiac components, as well as certain amount of motion artefacts and disturbances from the surrounding environment.

The impedances were measured using Zurich Instruments HF2IS Impedance Spectroscope. The classical Kelvin type four electrode configurations were used.

Current source excitation at slightly different frequencies was used in both of them to minimize mutual influence of simultaneous impedance measurements between two channels.

The measurement current was kept under 1mA at all times for safety reasons. Connection to the chest was made using Kendall/Tyco ARBO disposable surface EMG/ECG silver/silver chloride electrodes, followed by proprietary front end electronics close to the electrodes.

Biologically modulated impedance signals were collected from two orthogonal directions. One set of electrodes was placed on belt surrounding the chest, and another on vertical line between heart tip and neck.

The later analysis was accomplished in MATLAB environment on PC. The purpose of this study is to use the PCA and fast ICA individually or in combination in order to observe their capability to separate the cardiac and respiratory components of the total EBI. It is known that the spectra of the both components are overlapping each other, and thus it is a challenging task to separate these components in frequency and time domain.

The rest of this paper is structured as follows. Section 2 and 3 presents the related work of PCA and ICA. Section 4 discusses the block diagram and steps of the trial method. Section 5 demonstrates the results which are obtained. Finally, Section 6 concludes with the summary of the main contributions of the paper and future work.

## II. PRINCIPAL COMPONENT ANALYSIS (PCA)

PCA, which is sometimes also called the Karhunen-Loève transformation, is broadly used for the representation of high-dimensional data and is commonly used as a preprocessing step to a projection of high-dimensional data into a low-dimensional subspace [12]. As well the PCA is frequently used for data reduction in statistical pattern recognition and to visualize the comparisons between the biological samples, and for filtering out noise [6, 7].

The mixing matrix equation is:

$$Y = AS', \quad (1)$$

where $Y$ is data modelled as the product of $A$ (scores are the amount of artefact's variable for particular sample) and $S'$ (loading is define new coordinate, which has highest variation).

The principal component analysis projects the data into a new space and finds the principal components (PCs) like $s_1$, $s_2$, ....., $s_N$, which are uncorrelated and orthogonal [7]. The PCs can effectively extract the related information form the data [7], so that they carry the maximum amount of variance likely by $N$ linear transformed components [1, 6]. Each PC consists of one score and one loading and PCs are given by $s_i = w_i^A \cdot y$, where $y$ and $w_i$ are observation vector and $i$-th is the PCA weight vector, and $(.)^A$ represents the transposition [1, 6].

The computation for the $w_i$ can be achieved by making the use of covariance matrix $E\{xx^A\} = R_x$, where $E\{.\}$ is the expectation operator. $w_i$ are the eigenvectors of $R_x$ that relate to the $N$ largest eigenvalues of $R_x$. The first PC $s_1$ points in the direction where the input has the highest variance, and second PC $s_2$ is orthogonal to the first PC and points a direction of highest variance when the first project has been subtracted, and so on [1, 6].

## III. INDEPENDENT COMPONENT ANALYSIS (ICA)

The ICA algorithm is an implementation of the Blind Source Separation (BSS) approach and it is a valuable technique to find the independent components (ICs) of a multivariate random variable. These components are in the direction, which the element of the random variable has no dependency. The ICs are used to decrease effects of noise and artefacts of signals [7, 9]. Because of this, ICA becomes good application of the BSS.

In contrast to the correlation based transformation the PCA, the ICA also reduces higher-order statistical dependencies for non-Gaussian distributed signals [1]. In latest literature, it has been presented that the independent components (ICs) from the ICA were better in separating the different kinds of biological groups than principal components (PCs) from the PCA [7, 10, 11].

## IV. METHOD

In this study the well-known algorithms used such as the PCA and fast ICA, were selected to investigate their ability to extract the information about the two components of the EBI dataset corresponding to the cardiac and respiratory activities of a human. The cardiac and respiratory signals are correlated due to their nature [Ch. 8 in 13]. However they could be viewed as uncorrelated under assumption that the correlation is relatively weak to cause sufficient errors.

The complex EBI was measured on two channels. The obtained four real valued signals were used as input dataset for the selected methods.

The block diagram of trial method, which was used to investigate the separation of cardiac and respiratory components, is shown in the Figure 1.

The first part is the PCA algorithm, is intended to un-correlate the EBI dataset, whilst the second part is the ICA algorithm, which tries to separate the cardiac and respiratory components the total EBI.

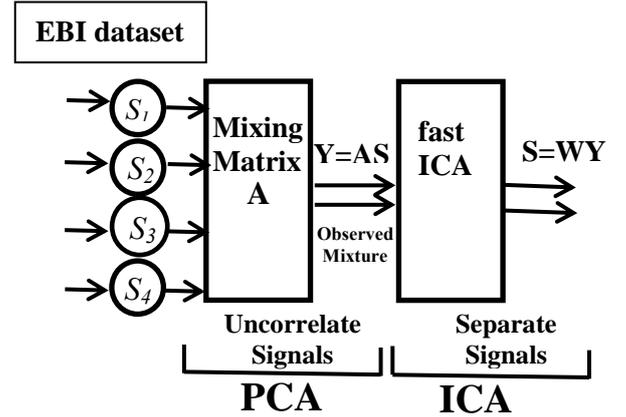

Figure 1. Block diagram of trial method, which consists the combination of PCA and ICA steps to solve the BSS problem.

It was observed that the PCA is capable to un-correlate and reduce the dimensionality of the EBI dataset but it is not efficient to separate the cardiac and respiratory signals accurately. However, ICA is more suitable for separation of independent components from a mixture of input signals [7, 10, 11]. The PCA and ICA algorithms complement each other since if only the PCA is used; no separation can be achieved, because the PCA only uncorrelates the data, it does not mean independence. However, if fast ICA is applied alone, it is too hard for fast ICA to solve problem.

The raw EBI signals were sampled at rate 1000 samples/s.

It was observed in this study that without preprocessing the raw data the ICA's convergence is slow.

The following steps were followed for solving the problem:
Step 1: The EBI dataset is loaded and divided into frames of 10,000 samples each.
Step 2: Frames of the EBI data are sampled down by factor 10.
Step 3: The second order low-pass Butterworth filter is applied to suppress the noise.
Step 4: The PCA is applied to un-correlate the filtered data.
Step 5: The fast ICA is applied to separate the cardiac and respiratory components.

Assume the EBI dataset is a centred $n * p$ matrix (the mean of each column has been subtracted), where $n$ is the number of samples (or observations) and $p$ is the number of variables or parameters that are measured.

$$Y = UDV^T, \qquad (2)$$

where $U$ is an $n * p$ matrix, columns of which are uncorrelated ($U^T U = I_p$), $D$ is a $p * p$ diagonal matrix with diagonal elements $d_j$ and the $V$ is a $p * p$ orthogonal matrix ($V^T V = I_p$).

After un-correlating the EBI dataset by the PCA method, the uncorrelated data $Y$ is passed as input to the fast ICA algorithm. The fast ICA algorithm maximizes the non-Gaussian form for each component and separates the independent data [5, 7].

Let $Y (n * p)$ be the centred data and $S (n * p)$ the matrix containing the independent components (ICs). It can solve the ICA problem by introducing a mixing matrix $A$ of size $n * n$.

$$Y = AS. \qquad (3)$$

The mixing matrix $A$ shows how the ICs of $S$ are linearly joined to make $Y$. If it reorder the equation above to get

$$S = WY, \qquad (4)$$

where the un-mixing matrix $W (n * n)$ describes the inverse process of mixing the ICs, if assuming $A$ is a square and orthonormal matrix and then the $W$ is basically the transpose of $A$. In practice, it is very beneficial to whiten the data matrix $Y$. In this study PCA is used as pre-processing step to centring and whitening the data matrix for fast ICA algorithm [5, 7]. However, fast ICA uses the PCA as a preprocessing by default.

After adopting the method, the convergence of the entire algorithm became faster and results are enhanced as well with regard to results obtained by applying the fast ICA algorithm alone as it is visible from Figure 2 (a).

## V. RESULTS

This study investigates the performance of both algorithms, the PCA and the fast ICA, to separate the EBI signals corresponding to cardiac and respiratory activities.

After some attempts and observations, it was understood that sequential use of both algorithms (PCA and ICA) are required in order to take advantage of BSS. The required steps are discussed in the method section IV.

In the Figure 2 the comparison of results achieved by applying both algorithms and described steps are depicted.

After following the steps described in method section, fast ICA performance was tested without the PCA. The results are depicted in the Figure 2 (a). However, fast ICA uses the PCA by default.

The results are depicted in the Figure 2 (b) for the case PCA separately was used as the preprocessing step for the fast ICA.

**Result of the fast ICA without using PCA as pre-processing**

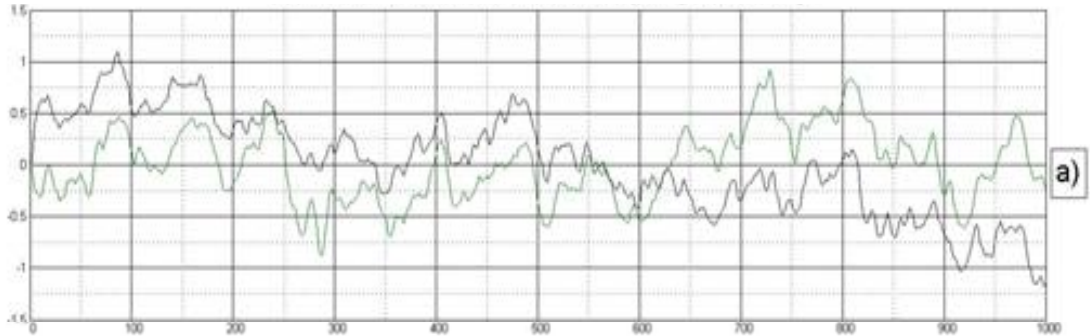

**Result of the fast ICA with using PCA as pre-processing**

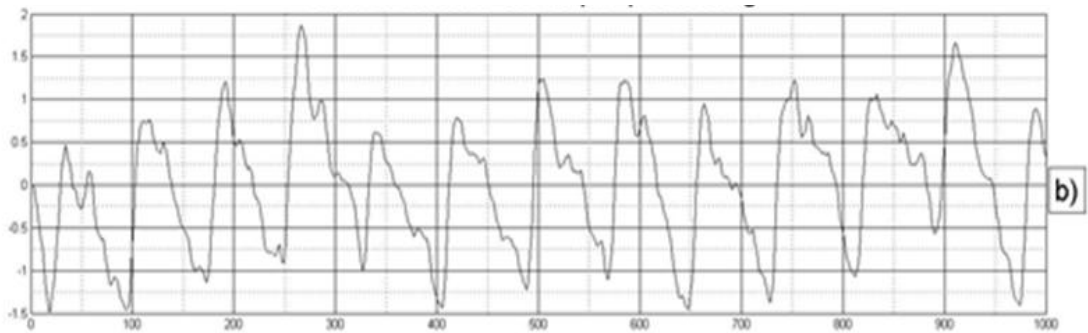

Figure 2. Results of investigation of the fast ICA applied on a frame of the EBI dataset:
(a) without using the PCA for pre-processing,
(b) with using the PCA for pre-processing.

Figure 2 (a) depicts the performance of fast ICA after applying only centring procedure to the EBI dataset. It is difficult to predict anything based on shown signals in Figure 2 (a) because signals are not clear. The results were discussed with cardiologist. The cardiologist said that based on figure 2 (a), it is difficult to perceive anything. However, based on figure 2 (b), could be some useful information.

On the other hand, it is observed that fast ICA alone is not efficient. The results are not so promising, which are depicted in Figure 2 (a).

In the second trial the PCA has been used as preprocessing step in order to reduce the dimension of the EBI dataset before applying the fast ICA algorithm. The more promising results were obtained, but not accurate enough to solve the task. The cardiac component still contains the respiratory one, but with lower amplitude.

However, the respiratory signal was not separated by both methods.

## VI. CONCLUSIONS AND FUTURE WORK

The complex EBI measured using two orthogonal placed electrode pairs. Thus four real valued signals were used as inputs for the PCA and ICA.

It was understood that the sequential use of both algorithms and proper steps of preprocessing is required to outcome the fast execution of algorithms; the steps are discussed in the method section. Firstly, it is required to reduce the demission of the EBI dataset, and then to separate the cardiac and respiratory signals.

The investigation showed that neither the PCA nor ICA nor combination of them can not accurately and robustly separate the components, at least when using only two complex valued (four real valued) input components. But the combination of the PCA and ICA algorithms showed more promising results, than the ICA alone, however the estimate of the cardiac component still contains the respiratory one.

The PCA uncorrelate and reduce the dimension of data. It does not mean the separation of components. The ICA works on the assumption of independence among source signals; if the source signals do not satisfy the condition then ICA would not be able to separate the components. In this case more investigation is required to understand the nature of cardiac and respiratory components.

It was the first attempt to use the PCA and fast ICA in order to separate cardiac and respiratory components from the electrical bio-impedance (EBI) dataset, measured using two orthogonal placed electrode pairs simultaneously.

Development of the method is required to approach better separation of cardiac and respiratory components from the EBI in order to take advantage of the BSS, machine learning techniques and understand the nature of cardiac and respiratory components.

ACKNOWLEDGMENT

The authors express thanks to Prof. T. Rang, Dr. T. Parve, Prof. M. Min, Dr. R. Land, and Dr. R. Gordon for providing valuable advices and practical information and others for the fast ICA package which is used in this study.

This research was supported by European Social Fund's Doctoral Studies, Internationalization Program DoRa, the Estonian Ministry of Education and Research (the target oriented project SF0140061s12), the Estonian Science Foundation (the research grants G8592 and G8905) and the Foundation Archimedes through the Center of Excellence CEBE (TK05U01).